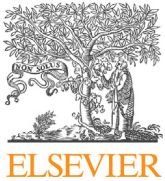
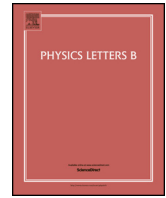

# Odd tensor electric transitions in high-spin Sn-isomers and generalized seniority

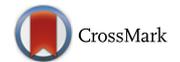

Bhoomika Maheshwari*, Ashok Kumar Jain

*Department of Physics, Indian Institute of Technology, Roorkee 247667, India*

**A B S T R A C T**

The similar behavior of the $B(E1)$ values of the recently observed $13^-$ odd tensor $E1$ isomers and the $B(E2)$ values of the $10^+$ and $15^-$ even tensor $E2$ isomers in the Sn-isotopes has been understood in terms of the generalized seniority for multi-$j$ orbits by using the quasi-spin scheme. This simple approach proves to be quite successful in explaining the measured transition probabilities and the corresponding half-lives in the high-spin isomers of the semi-magic Sn-isotopes. Hence, we show for the first time the occurrence of seniority isomers in the $13^-$ Sn-isomers, which decay by odd-tensor $E1$ transitions to the same seniority states.

© 2015 The Authors. Published by Elsevier B.V. This is an open access article under the CC BY license (http://creativecommons.org/licenses/by/4.0/). Funded by SCOAP$^3$.

## 1. Introduction

Recent years have seen a resurgence in the discovery of new isomers by using modern experimental techniques. It is, therefore, now possible to identify not so common situations where seniority isomers may occur. Many experimental groups [1,2] have recently reported the transition probability measurements of the seniority $v = 4$, $15^-$ and $13^-$ isomers in the Sn-isotopes. Iskra et al. [2], in particular, have recently highlighted the surprisingly similar trends of the experimental $B(E2)$ values in the $15^-$ isomers and the $B(E1)$ values in the $13^-$ isomers for the even–even Sn-isotopes, and have gone to the extent of questioning their own measurements due to the totally unexpected regular behavior of the $B(E1)$ values in the $13^-$ isomers.

The seniority scheme, first introduced by Racah [3], has widely been used to understand the semi-magic nuclei, particularly where a single-$j$ orbital plays the dominant role [4–6]. The behavior of the odd/even tensors in the pure seniority scheme for the single-$j$ shell has been studied [4,5] by using the quasi-spin formalism of Kerman [7,8] and Helmers [9]. It is now known that the odd tensor single-particle operators are quasi-spin scalars, while the even tensors become the $\sigma = 0$ components of the quasi-spin vector [4]. As a result, the odd tensor transitions conserve the seniority for both electric and magnetic case, and lead to a particle number independent behavior. On the other hand, the even tensor electric or magnetic transitions may change the seniority by 2. It is also well known that the even tensor transition probabilities between the same seniority states ($v \to v$) vanish at the mid-shell displaying a parabolic behavior, and lead to seniority isomers. The large body of data on the seniority isomers has so far, therefore, been based on the $E2$ transitions alone [5]. We note that favorable conditions do not exist for the occurrence of seniority isomers involving parity changing odd tensor electric transitions as a parity change is not possible in the case of single-$j$ orbit. Arima and Ichimura [10] were the first to introduce the concept of generalized seniority in the case of multi-$j$ orbits, mainly for the degenerate orbits. Talmi further extended the concept of generalized seniority to the case of several non-degenerate orbits and the resulting mixed configuration states were used to understand the semi-magic nuclei in the shell model [11,12].

In the present letter, we extend the usage of the quasi-spin formalism for generalized seniority [13] to calculate the reduced transition probabilities of odd as well as even multipole electric transitions in multi-$j$ degenerate orbits. By using these results, we are able to reproduce remarkably well, the similar behavior of the measured half-lives and transition probabilities of $E1$ and $E2$ transitions for the high-spin $13^-$ and $10^+$, $15^-$ isomers in the semi-magic $^{116-130}$Sn isotopes. We are thus able to confirm for the first time the occurrence of seniority isomers, which arise from the odd tensor electric transitions, and in particular $E1$ transitions.

We also note that the $10^+$ isomers involving the $E2$ transitions have been understood in the literature as pure seniority $v = 2$

* Corresponding author.
*E-mail address:* bhoomika.physics@gmail.com (B. Maheshwari).

http://dx.doi.org/10.1016/j.physletb.2015.11.079
0370-2693/© 2015 The Authors. Published by Elsevier B.V. This is an open access article under the CC BY license (http://creativecommons.org/licenses/by/4.0/). Funded by SCOAP$^3$.

states, and the parabolic nature of the experimental $B(E2, 10^+ \to 8^+)$ values with a minimum at the mid-shell has been qualitatively explained in terms of the single-$j$ $h_{11/2}$ configuration [4,14,15]. While theoretical calculations for $B(E2, 0^+ \to 2^+)$ values in Sn-isotopes have recently been reported by Morales et al. [16] in terms of the generalized seniority scheme, no quantitative results have so far been reported for the $B(E2)$ values of the high-spin isomers such as the $10^+$ isomers. We show that it is necessary to invoke the concept of generalized seniority to fully explain the behavior of the $E2$ transition probabilities in the well known examples of the $10^+$ isomers in Sn-isotopes also, which have so far been understood as pure seniority isomers.

From our calculations, we are thus able to explain the similar behavior of the odd tensor electric transition probabilities and the even tensor electric transition probabilities between the states having the same generalized seniority in multi-$j$ configurations. This simple approach may prove useful for estimating the transition probabilities in those cases for which measurements have not been made so far.

## 2. Formalism used

In the quasi-spin scheme of Kerman [7], with multi-$j$ degenerate orbits, the commutator between the correlated pair creation operator $S^+$ and a single particle hermitian tensor operator $T_\kappa^{(k)}$ is given by [4];

$$[S^+, T_\kappa^{(k)}] = \frac{1}{\sqrt{2k+1}} \sum_{j<j'} [1 + (-1)^k](j||T^{(k)}||j')(a_j^+ \times a_{j'}^+)_\kappa^{(k)} \quad (1)$$

where $S^+ = \sum_j S_j^+$ is the sum of the pair creation operators for different-$j$ orbits. The equation (1) implies that if $k$ is odd, then the Hermitian operator becomes a quasi-spin scalar, which is a simple generalization of the case when $j = j'$. However, if $k$ is even then $T_\kappa^{(k)}$ becomes the $\sigma = 0$ component of a quasi-spin vector, again a generalization of the case $j = j'$, and follows the well known algebra accordingly, as shown in Chapter 19 of Talmi's book [4].

Arvieu and Moszokowski [13] further generalized the relations for the multi-$j$ configuration by using the quasi-spin operators defined by $S_1^+ = \sum_j (-1)^{l_j} S_j^+$, where $l_j$ is the orbital angular momentum of the $j$-orbit. The phase factor appearing in Eq. (1) then becomes $(-1)^{l+l'+L}$, where $L$ represents the nature of the tensor involved in the transition between the initial and final states whose parities are governed by the $l$ and $l'$ values. Therefore, when $(-1)^{l+l'+L}$ is equal to $-1$, the Hermitian operator becomes a quasi-spin scalar, and conserves the seniority in multi-$j$ configurations as well, which is the case of the magnetic transitions. On the other hand, the commutator between $Y_M^L$ and $S_1^+$ does not vanish for any $L$ value, even or odd, for electric transitions, when $(-1)^{l+l'+L}$ is equal to $+1$. Thus, the electric multipole operators, even or odd, become the $\sigma = 0$ components of the quasi-spin vector in the multi-$j$ case and follow the same algebra. We can easily rewrite the relations of the single-$j$ case for the multi-$j$ case by defining $\tilde{j} = j \otimes j' \ldots$, with the corresponding total pair degeneracy $\Omega = \frac{1}{2}(2\tilde{j} + 1) = \frac{1}{2}\sum_j(2j+1)$. One may, therefore, calculate the reduced transition probabilities for $n$ particles in multi-$j$ configuration as,

$$B(EL) = \frac{1}{2J_i+1} |\langle \tilde{j}^n v l J_f || \sum_i r_i^L Y^L(\theta_i, \phi_i) || \tilde{j}^n v' l' J_i \rangle|^2 \quad (2)$$

The reduced matrix elements can be written in terms of the seniority reduction formulae within the multi-$j$ configuration for $\Delta v = 0$ and $\Delta v = 2$, by using the same quasi-spin algebra as applicable in the single-$j$ case [4], so that

$$\langle \tilde{j}^n v l J_f || \sum_i r_i^L Y^L || \tilde{j}^n v l' J_i \rangle$$
$$= \left[ \frac{\Omega - n}{\Omega - v} \right] \langle \tilde{j}^v v l J_f || \sum_i r_i^L Y^L || \tilde{j}^v v l' J_i \rangle \quad (3)$$

$$\langle \tilde{j}^n v l J_f || \sum_i r_i^L Y^L || \tilde{j}^n v \pm 2 l' J_i \rangle$$
$$= \left[ \sqrt{\frac{(n-v+2)(2\Omega+2-n-v)}{4(\Omega+1-v)}} \right]$$
$$\times \langle \tilde{j}^v v l J_f || \sum_i r_i^L Y^L || \tilde{j}^v v \pm 2 l' J_i \rangle \quad (4)$$

We note that the coefficients in the square brackets, which arise for the multi-$j$ orbits in the present case, are identical to the well known results for single-$j$, due to the simple extension to the many-$j$ problem by defining $\tilde{j} = j \otimes j' \ldots$. However, the interpretation of the results changes due to the extra phase factor arising in the multi-$j$ case. The electric multipole transition probabilities $B(EL)$ show a minimum at the middle of the multi-$j$ shell, for transitions between the same seniority states. On the other hand, they show a peak at the mid-shell for transitions between the states with a change in seniority by 2. The behavior of the $B(EL)$ values, thus, becomes independent of the nature of the tensor $L$, whether even or odd, in the multi-$j$ situation. Therefore, the odd tensor electric transition probabilities also behave similarly to the even tensor electric transitions, and this may also lead to seniority isomers. In the special case of the surface delta interaction (SDI), the radial part of the matrix elements may be taken as approximately constant [13,17], and the $B(EL)$ values simply vary according to the coefficients which depend on $\Omega$, $n$ and $v$.

## 3. Calculations and results

We apply these results to explain the recently measured similar trends of the $B(E1)$ values in the $13^-$ isomers and the $B(E2)$ values in the $10^+$ and $15^-$ isomers in the Sn-isotopes. Assuming $\Delta v = 0$ in the decay of all the high-spin $10^+$, $13^-$ and $15^-$ isomers, the $B(EL)$ should be proportional to the coefficient $((\Omega - n)/(\Omega - v))^2$. We calculate the variation of the $B(EL)$ values in the $^{116-130}$Sn isotopes, where measurements have been made, as follows.

We assume that the $g_{7/2}$ and $d_{5/2}$ orbits are completely filled in $^{114}$Sn, which is taken as a core; hence, the active orbits for these isomers are $h_{11/2}$, $d_{3/2}$ and $s_{1/2}$ in the 50–82 valence space. Since these orbits have small relative energy splittings compared to the usual pairing gap, we may treat them as approximately degenerate. A combination of the three orbits can provide several mixed configurations having resultant $\tilde{j}$ and $\Omega$ values as: $\tilde{j} = h_{11/2} \otimes d_{3/2} \otimes s_{1/2}$, $\Omega = 9$; $\tilde{j} = h_{11/2} \otimes d_{3/2}$, $\Omega = 8$; $\tilde{j} = h_{11/2} \otimes s_{1/2}$, $\Omega = 7$; and $\tilde{j} = j = h_{11/2}$, $\Omega = 6$.

For all the isomeric transitions discussed in this paper, we find that only $\Omega = 9$ and $\tilde{j} = h_{11/2} \otimes d_{3/2} \otimes s_{1/2}$ is able to reproduce the measured values reasonably well. This implies that the $10^+$, $13^-$ and $15^-$ isomers follow the generalized seniority scheme, and the configuration mixing is essential. The numerical results are obtained by fixing the proportionality constant using the first available experimental value (on the extreme left in Figs. 1 and 2) of the transition probability. The proportionality constants in the $10^+$ isomers remain constant at 54 for all the possible omega values, since the coefficient $\left((\Omega - n)/(\Omega - v)\right)^2$ becomes 1 as $n = v = 2$ for all $\Omega$ values. On the other hand, these constants become omega-dependent for the $13^-$, and $15^-$ isomers. The proportionality constants in the $13^-$ isomers are found to be 12.77, 18.4 and 41.4 for $\Omega = 9, 8$ and 7, respectively, whereas these constants in the $15^-$ isomers are obtained as 58.33 and 84 for $\Omega = 9$ and 8, respectively. Note that these proportionality constants contain the information on the radial integrals along with the reduced matrix elements of the spherical harmonics $Y_M^L$ in the multi-$j$ case.

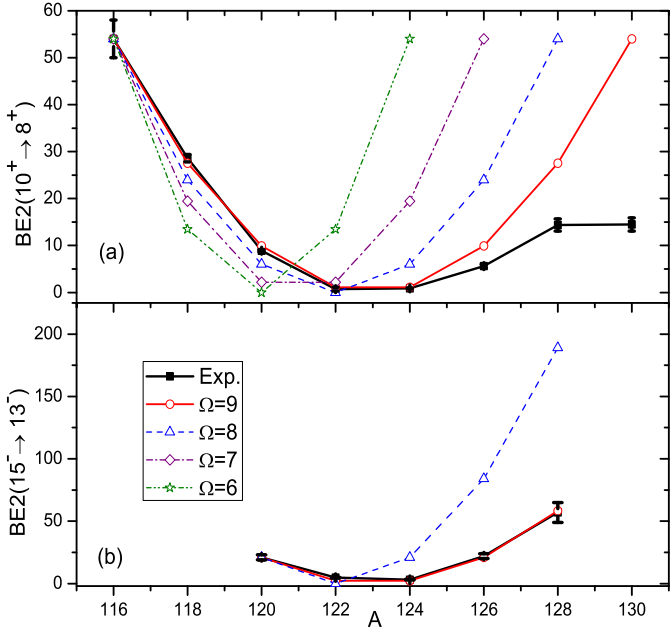

**Fig. 1.** (Color online.) Variation of the $B(E2)$ values (in the units of $e^2\,\text{fm}^4$) for the $10^+$, and $15^-$, $E2$ isomers in Sn-isotopes. Experimental data are taken from [2,14, 15].

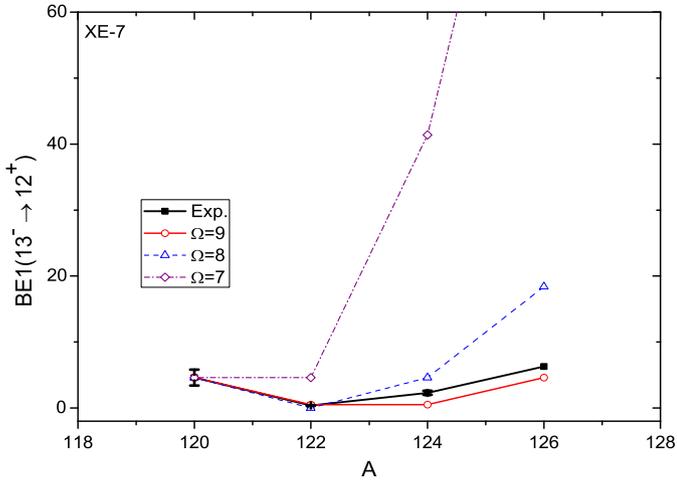

**Fig. 2.** (Color online.) Variation of the $B(E1)$ values for the $13^-$, $E1$ isomers in Sn-isotopes (in the units of $e^2\,\text{fm}^2$). Note that the vertical scale has a multiplier of E-7, as mentioned.

The $10^+$ isomers in the Sn-isotopes have generally been understood in the literature as arising from the pure $h_{11/2}$, seniority $v = 2$ configurations. The parabolic behavior of the $B(E2)$ values has also been qualitatively understood in terms of the seniority scheme. In Fig. 1(a), we have calculated the $B(E2)$ (in the units of $e^2\,\text{fm}^4$) for $10^+$ isomers in the $^{116-130}$Sn isotopes with several possible omegas (follow the color coding). Experimental data [14, 15] are shown by the filled square symbols. We find that the experimental data are reproduced by the calculations quite well only for $\Omega = 9$ arising from $\tilde{j} = h_{11/2} \otimes d_{3/2} \otimes s_{1/2}$, except for a deviation in the $^{128,130}$Sn isotopes where the values appear to have become constant. We, therefore, conclude that all the three orbits are necessary to explain the behavior of the $10^+$ isomers, and the generalized seniority needs to be invoked, which still remains constant at $v = 2$ for these isotopes. Note that $\Omega = 6$ corresponds to the pure $h_{11/2}$ configuration; while it does lead to a parabolic behavior, it is quite far from the measured values. The constant value of $B(E2)$s for $^{128,130}$Sn is most probably due to the constant occupancy at $N = 78$ and 80 for the $h_{11/2}$ orbit, which is the dominant orbit.

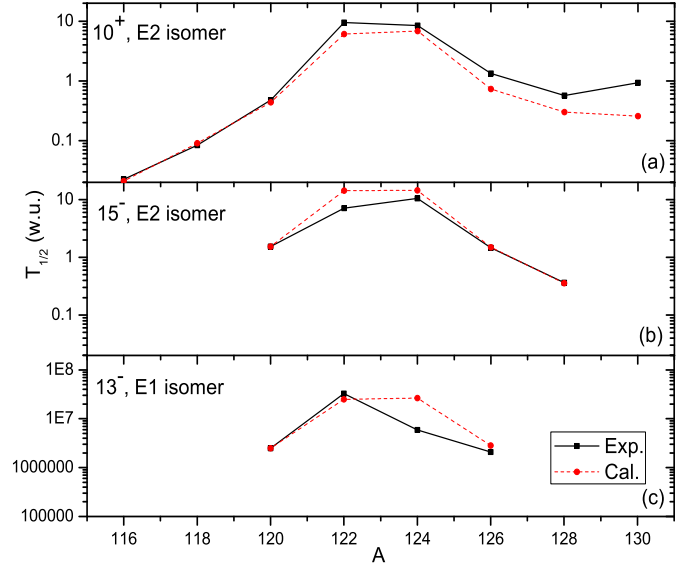

**Fig. 3.** (Color online.) Variation of the half-lives of the $10^+$, $15^-$ and $13^-$ isomers in Sn-isotopes. Experimental half-lives have been taken from the atlas [18]. All the values are shown in Weisskopf units (w.u.).

Similarly, we have calculated the reduced transition probabilities for the higher seniority $v = 4$, $13^-$ and $15^-$ isomers for various $\tilde{j}$ and omega values (see Figs. 1(b) and 2). Experimental data are taken from the recent work of Iskra et al. [2]. We again find that the experimental data are reproduced quite well, when the configuration mixing is $\tilde{j} = h_{11/2} \otimes d_{3/2} \otimes s_{1/2}$ with $\Omega = 9$. Iskra et al. [2] have reported the configurations of the $13^-$ and $15^-$ isomers as $h_{11/2}^3 \otimes s_{1/2}$, and $h_{11/2}^3 \otimes d_{3/2}$ (seniority $v = 4$), respectively. However, we find that the involvement of all the three active orbits ($\Omega = 9$) is necessary to explain the measured trends of the transition probabilities in both the set of isomers (see Figs. 1(b) and 2); the other omega values yield results far from the measurements. We, therefore, confirm that generalized seniority still remains constant at $v = 4$ and the decay of these isomers corresponds to $\Delta v = 0$. As expected, the excitation energies of the $13^-$ and $15^-$ isomers relative to the $10^+$ isomers also remain nearly constant for $^{120-128}$Sn isotopes [2,18].

As $\Omega = 9$ alone has been able to explain the measured systematics for all the three isomers, we have calculated the half-lives of these isomers by using the calculated transition probabilities for $\Omega = 9$, the internal conversion coefficients [19], and the experimental gamma ray energies, intensities and branching ratios [2,14,15,18,20]. Fig. 3 presents a comparison of the experimental and the calculated half-lives in Weisskopf units (w.u.) for all the three sets of isomers in the three separate panels. The calcu-lated trends reproduce the experimental half-lives quite well (see Fig. 2). The peaks at the middle are very obvious due to the fact that the transition probabilities show minima at the same mass numbers, irrespective of the nature of the involved electric tensors, which may be even or odd. Therefore, the significant rise in the half-lives, particularly at the middle, is due to the role of seniority.

## 4. Conclusion

To conclude, we have used the quasi-spin formalism for degenerate multi-$j$ orbits to calculate the reduced electric transition probabilities in the semi-magic isomers, for both the even and odd tensor transitions. It has generally been believed that only $E2$ transitions can lead to the seniority isomers, an obvious conclusion from the seniority formalism in the case of single-$j$ orbit. We have shown that both the odd and even tensors behave similar to each other when many-j orbits are involved. As a result, the odd tensor transitions may also lead to seniority isomers. We have applied these results to the recent measurements in the $10^+$, $13^-$ and $15^-$ isomers in $^{116-130}$Sn isotopes, and successfully showed that the trends of $B(E1)$ values in the $13^-$ isomers and $B(E2)$ values in the $10^+$, $15^-$ isomers are very similar to each other. Hence, for the very first time, the seniority isomers with odd tensor electric transitions have been identified and interpreted in terms of the generalized seniority. We also find that the configuration mixing is essential to fully describe the even tensor transitions for the $10^+$ isomers also, which were so far interpreted as pure $h_{11/2}$ isomers. This simple scheme of calculating the $B(EL)$ values may also be used to estimate the half-lives in unknown cases and, hence predict new isomers. This scheme may be further generalized for the non-degenerate orbits to obtain a more complete description within the Talmi's generalized seniority scheme [11] along the lines of Morales et al. [16].

## Acknowledgements

We would like to thank P. Van Isacker and R.F. Casten for a critical reading of the manuscript and valuable discussions. Financial support from the Department of Science and Technology, Department of Atomic Energy, and Ministry of Human Resource Development (Government of India) is gratefully acknowledged.